\documentclass[aps,pre,reprint,groupedaddress]{revtex4-1}
\usepackage{graphicx}
\usepackage{amsmath,amssymb,amsfonts}

\newcommand{\be}{\begin{equation}}
\newcommand{\ee}{\end{equation}}
\newcommand{\bea}{\begin{eqnarray}}
\newcommand{\eea}{\end{eqnarray}}

\def\d{\delta}
\def\p{\phi}
\def\a{\alpha}
\def\b{\beta}
\def\l{\lambda}
\def\c{\chi}
\def\s{\sigma}
\def\eps{\epsilon}

\def\sst{\scriptscriptstyle}

\begin{document}


\title{Information-theory-based solution of the inverse problem in classical statistical
mechanics}
\thanks{http://link.aps.org/doi/10.1103/PhysRevE.82.021128}

\author{Marco D'Alessandro}
\email[e-mail address: ]{marco.dalessandro@isc.cnr.it}
\author{Francesco Cilloco}

\affiliation{Institute for Complex Systems, National Research Council (CNR), Via del
Fosso del Cavaliere 100, 00133 Rome, Italy}

\begin{abstract}
We present a procedure for the determination of the interaction potential from the knowledge
of the radial pair distribution function. The method, realized inside an inverse Monte Carlo
simulation scheme, is based on the application of the maximum entropy principle of information
theory and the interaction potential emerges as the asymptotic expression of the transition
probability.
Results obtained for high density monoatomic fluids are very satisfactory and provide an accurate
extraction of the potential, despite a modest computational effort.
\begin{description}
\item[DOI] 10.1103/PhysRevE.82.021128
\item[PACS numbers] 05.20-y, 89.70.Cf, 61.20.Ja
\end{description}
\end{abstract}

\maketitle


\section{\label{Introduction}Introduction}

This paper deals with the `inverse problem' in classical statistical mechanics.
Namely we are interested in determining the interaction potential of a system from
the knowledge of its radial distribution function (RDF).
A basic result due to Henderson \cite{Henderson1974} states that if a system is
governed by pairwise additive interactions then two potentials which give rise to
the same RDF cannot be different more than a constant term.
This theorem provides a theoretical support to the formulation of the
inverse problem since it demonstrates the uniqueness of its solution.
However the existence of the solution is not guaranteed and furthermore the
theorem does not indicate a way to find it.

Despite this general result, the solution of the inverse problem for a classical
dense fluid turns out to be a difficult task to achieve. This is due mainly to the
fact that in the high density regime the RDF is hardly sensitive to the detailed
shape of the interaction potential and is essentially determined by its repulsive part;
so the inverse functional relationship between the RDF and the interaction potential
evidences a strong dependence of the latter on the input RDF.
In order to expect a reliable solution of the inverse problem not only the input
RDF must be provided with high precision but also the underlying theory used to
formulate the inversion procedure must be very accurate.
As stated by Reatto in \cite{Reatto1988} the accuracy of a satisfactory inversion
scheme has be to independent both from the shape of the interaction potential and
from the density of the system under inspection. If these properties are fulfilled
then the interaction potentials of different systems can be consistently compared,
furthermore any dependencies of the extracted potential on the thermodynamic state can
be unambiguously ascribed to the effects of many-body interactions.

A generally accepted scheme for the solution of the inverse problem which fulfills
these features is still lacking and, in the last three decades, several authors have
proposed different approaches.
A first category comprises theoretically based attempts in which the inversion
scheme is defined on the basis of an integral equation theory (HNC, MHNC, etc.)
of the liquid state. These pure theoretical approaches typically rely on some
approximation and, due to the intrinsic difficulties depicted above, their application
provides reliable results only in a limited set of cases.
A considerable improvement in the accuracy of the extracted potentials has been
obtained by recurring to simulation assisted procedures.
These methods attempt to determine the pair interaction starting from a
guessed expression of the potential which is iteratively modified on the basis of the
discrepancy between the simulated pair function and the experimental data.
A first result in this direction has been proposed by Schommers in \cite{Schommers1983}
and later on further improvements have been achieved by Reatto, Levesque and Weiss in
\cite{Reatto1986}; in this paper the authors applied the predictor-corrector scheme,
using the MHNC equation as predictor, to the Lennard-Jones fluid and to a model potential
for aluminum. The convergence of the iterative potential to the correct result was found
and it was checked that the use of a less accurate predictor (for example the one proposed
in \cite{Schommers1983}) for the definition of the trial potential could spoil the accuracy
of the procedure.
Other results belonging to this class of inversion procedures comprise the empiric
 potential structure refinement (EPSR) proposed by Soper \cite{Soper1996,Soper2001} and a solution
due to Lyubartsev and Laaksonen \cite{Lyubartsev1995}. The former technique performs the refinement
of a reference potential using a perturbation term given by the difference between the experimental
and the simulated structure factor; the latter propose a parametric dependence of the potential on
a set of parameters which are determined by solving a large system of linear equations.

A further approach to the inverse problem is provided by a family of `stochastic'
inversion methods in which the solution is sought as the expected value of
properly extracted random variables (inverse Monte Carlo).
In this simulation scheme, given the input RDF, a dynamical evolution law is defined
with the aim to build a set of configurations compatible with the experimental data.
So the solution of the inverse problem is brought back to the determination of a suitable
transition probability which produces a `Monte Carlo like' dynamic.
Among the various attempts in this direction we mention the Reverse Monte Carlo
(RMC) technique due to McGreevy and Pusztai \cite{McGreevy1988} and two `absolute
minimization' methods proposed by Cilloco in \cite{Cilloco1993} and later on by
da Silva, Svensson, \AA kesson and J\"{o}nsson in \cite{daSilva1998}.
Strictly speaking these methods do not provide a solution of the inverse problem
since they do not allow the direct determination of the interaction potential,
however the configurations produced in the inverse Monte Carlo procedure can be
used to compute quantity of physical interest.
It is worth mentioning that results reported in \cite{Cilloco1993} represent the first
application of the maximum entropy principle, indicating a possible solution based on the measurement
of the three body correlation function.
A further contribution is due to a technique proposed by Almarza, Lomba and Molina \cite{Almarza2003,Almarza2004}
where a direct solution of the inverse problem has been obtained by performing a continuum
refinement procedure of a trial interaction potential.

The purpose of this paper is to present a technique for the solution of the inverse problem
based on the maximum entropy principle (ME) \cite{Jaynes1957}.
ME is an effective tool for setting up the equilibrium distribution of a statistical system
on the basis of partial knowledge and the corresponding estimate fulfills the remarkable
property of being the `maximally noncommittal with regard to the missing information'.
So, our solution of the inverse problem is based on the maximization of the configurational
entropy constrained by the information codified in the radial pair distribution function.
The procedure is realized inside an inverse Monte Carlo scheme and the interaction potential
emerges as the asymptotic expression of the transition probability.

The contents of the paper are as follows. Section \ref{Theory} contains a description
of our method, in section \ref{Applications} we test the method in the case of a
Lennard-Jones fluid and for a model of liquid aluminum. Finally in section \ref{Discussion}
we discuss our results and present some final remarks.

\section{\label{Theory}Theory}

\subsection{\label{Statistical}Statistical description of a monoatomic system}

We perform a statistical analysis of a simple monoatomic system with the aim to define
some quantities that will be of central interest later on in the paper. Particular
emphasis will be given to the concepts of probability, likelihood, entropy and to
their mutual relationship.

Consider an homogenous and isotropic system composed of point-like elements
with average density $\rho$. In the following we will refer to this system as the \emph{model}.
Given an arbitrary configuration $\bf{x}$ of the model we can perform a local sampling
of the elements pair function (PF). This means that we select a reference element
and divide the space in spherical shells of width $\d r$ centred on it up to the maximum
value $r_{M}$ \footnote{the configuration space of the system is assumed large enough so
that for each of the sampled elements the sphere of radius $r_{M}$ does not cross the boundary
of the system.}; then we count the number of elements in each shell and we store these
numbers in the array $n_{i}$, where $i = 1,...,J$.

We define a probability function $p(\bf{x})$ over the configuration space of the model
system and collect an ensemble of $s$ configurations extracted according to $p(\bf{x})$.
The global sampling of the PF over the ensemble can be computed by evaluating the $n_{i}$
for each configuration $\a$ and summing these local samplings shell by shell, that is:
\be \label{mdef1}
m_{i} \doteq \sum_{\a = 1}^{s} n_{i}^{(\a)}
\ee
Assuming that the expected PF is given by a reference function $\mu_{i}$ we can evaluate
the probability associated to the global sampling \eqref{mdef1}.
Let us focus on a fixed shell $k$. The values of $n_{k}$ obtained in two different
configurations are uncorrelated and, admitting that the number of shells is large
enough, the probability of finding more than one element in a single measurement can be
neglected; so the shell $k$ follows a Poisson distribution with expected value $s\mu_{k}$.
Since there is no correlation between different shells we obtain:
\be \label{poissondef1}
{\cal P}_{\mu}(m) = \prod_{i=1}^{J} e^{-s\mu_{i}}
\frac{(s\mu_{i})^{m_{i}}}{m_{i}!}
\ee
so the probability associated to $m$ is given by a product of Poisson distributions.
This formula describes an `open system', which can be realized as an open subset of a larger
one, and elements fluctuations are possible.
Conversely, if we are dealing with a closed system in which each configuration of the ensemble
satisfies the further constraint:
\be \label{multinomconstraint1}
\sum_{i=1}^{J} n_{i} = \sum_{i=1}^{J} \mu_{i} = N_{p}
\ee
the total number of elements is conserved and the probability \eqref{poissondef1} is reduced
to a multinomial expression (see \cite{Baker1984} and references therein):
\be \label{multinomdef1}
{\cal M}_{q}(m) = \frac{{\cal P}_{\mu}(m)}{P \left(\sum_{i} n_{i} = N_{p} \right)} =
N! \prod_{i=1}^{J} \frac{q_{i}^{m_{i}}}{m_{i}!}
\ee
where $N \doteq sN_{p}$ and $q_{i}=\mu_{i}/N_{p}$ is the normalized reference probability
distribution. Equations \eqref{poissondef1} and \eqref{multinomdef1} can be interpreted as
likelihood functions ${\cal L}(m,\mu)$ of the expected PF given the observed values $m$.
If the number of configurations in the ensemble is very large ($s\gg 1$ which implies
$N,m_{i} \gg 1$) we can take the logarithm of ${\cal L}$ and make use of the Stirling
approximation up to the linear order, this gives:
\bea \label{logprobability1}
&& \ln {\cal L}_{\cal P}(m,\mu) \simeq
-\sum_{i} m_{i} \ln \frac{m_{i}}{s\mu_{i}} + \sum_{i} (m_{i}-s\mu_{i})
\nonumber \\
&& \ln {\cal L}_{\cal M}(m,\mu) \simeq
-\sum_{i} m_{i} \ln \frac{m_{i}}{N q_{i}}
\eea
the two formula in \eqref{logprobability1} differ for a linear term which accounts
for the fluctuations of elements.

The log-likelihood equations \eqref{logprobability1} possess a nice interpretation when the
number of configurations becomes infinite. Let us focus on the multinomial likelihood
given by the second line of \eqref{logprobability1}; in the asymptotic limit the average PF
converges to the probability $p_{i} = m_{i}/N$ built over the ensemble and the likelihood
can be written as:
\be \label{logprobability2}
\lim_{s \rightarrow \infty} \frac{1}{s} \ln {\cal L}_{\cal M}(m,\mu) \simeq
-N_{p} \sum_{i} p_{i} \ln \frac{p_{i}}{q_{i}} = - N_{p} D(p||q)
\ee
we recognize that the log-likelihood is proportional to the relative entropy
$D(p||q)$ (Kullback-Leibler divergence \cite{Kullback1951}) of the ensemble distribution $p$
with respect to the reference one.
The relative entropy fulfills the properties of being positive definite and vanishing only
if $p=q$. Equation \eqref{logprobability2} implies that if the global PF built over the model
ensemble maximizes the likelihood with the reference function $\mu$ then, asymptotically, the
distribution $p$ minimizes the relative entropy respect to the reference probability $q$.
We will make us of this property in the next section.

It is useful to rewrite equation \eqref{logprobability2} in term of radial distribution
functions. The model RDF $g(r_{i})$ and its reference counterpart $g_{\sst{0}}(r_{i})$ are
defined by normalizing the ensemble average and the expected reference function $\mu$ by
the average value of particle per shell, respectively. So we have:
\be \label{RDFdef1}
g(r_{i}) = \lim_{s \rightarrow \infty}
\frac{1}{s} \frac{m_{i}}{4\pi\rho r_{i}^{2} \d r} \qquad
g_{\sst{0}}(r_{i}) = \frac{\mu_{i}}{4\pi\rho r_{i}^{2} \d r}
\ee
Plugging equation \eqref{RDFdef1} in \eqref{logprobability2} and passing to the
continuum limit provides an expression for the relative entropy that will be widely
used in the following:
\bea \label{relativentropy1}
K_{\cal M}(g||g_{\sst{0}}) & = & \lim_{s \rightarrow \infty }
\frac{1}{2s} \ln {\cal L}_{\cal M}(m,\mu)\simeq \nonumber \\
& = & - \frac{\rho}{2} \int d {\bf r} g(r) \ln \frac{g(r)}{g_{\sst{0}}(r)}
\eea
where the extra factor $\frac{1}{2}$ has been inserted to avoid a double counting of the
number of independent distances between pairs of elements.
The same analysis can be repeated starting from the first line of equation
\eqref{logprobability1}; performing the asymptotic limit and recasting the
result in term of RDFs gives:
\bea \label{relativentropy2}
&&K_{\cal P}(g||g_{\sst{0}}) = \lim_{s \rightarrow \infty }
\frac{1}{2s} \ln {\cal L}_{\cal P}(m,\mu)\simeq \nonumber \\
&-& \frac{\rho}{2} \int d {\bf r} \left[ g(r) \ln \frac{g(r)}{g_{\sst{0}}(r)} -
(g(r) - g_{\sst{0}}(r)) \right]
\eea
which provides the relative entropy between the RDFs when elements fluctuations
are taken into account.

A last comment regards the meaning of this construction when a uniform reference
distribution $q_{i}=1/J$ is employed.
In this case equation \eqref{multinomdef1} provides the number of occurrences of the global
PF \eqref{mdef1} up to a constant factor and the relative entropy $D(p||q)$ reduces to the
Shannon entropy \cite{Shannon1948} up to an additive constant. Expressing this condition in
terms of RDFs supplies the measurement of the relative entropies \eqref{relativentropy1} and
\eqref{relativentropy2} respect to the `non-informative' reference system $g_{\sst{0}} \equiv 1$:
\be \label{entropytwodoby1}
S_{\cal M}^{\sst{(2)}} = K_{\cal M}(g||1)
\qquad
S_{\cal P}^{\sst{(2)}} = K_{\cal P}(g||1)
\ee
exploiting equations \eqref{relativentropy1} and \eqref{relativentropy2} we recognize
that the entropies \eqref{entropytwodoby1} exactly reproduce the two-body contribution
to the Boltzmann entropy expansion in the canonical ensemble \cite{Green1952} and in
the grand canonical ensemble \cite{Nettleton1958,Raveche1971}, respectively.

\subsection{\label{ME}Maximum entropy solution of the inverse problem}

We consider a monoatomic system whose interactions are governed by a
genuine pairwise additive potential $\p(r)$ and assume that for a given
condition of temperature $T$ and density $\rho$ the RDF of the system
$g_{\sst{0}}(r)$ is known. We refer to this system as the \emph{target}.
The interaction potential of the system is supposed to be unknown, only the RDF is
given.

We propose a solution of the inverse problem based on the maximum entropy principle
\cite{Jaynes1957} constrained by the information encoded in the RDF of the target system.
Namely we build a probability distribution $p$ in the model system which fulfills the
properties of maximizing the Shannon entropy consistently with the condition of vanishing
relative entropy with respect to $g_{\sst{0}}(r)$:
\be \label{relativentropy3}
K(g||g_{\sst{0}}) = 0
\ee
where the model RDF $g(r)$ is obtained by averaging the global PF \eqref{mdef1}
over an ensemble of configuration extracted according to $p$.
Formally this task is achieved by computing the maximum of the functional:
\be \label{Maxentropy1}
{\cal F}\{p\} = S\{p\} +\a K(g\{p\}||g_{\sst{0}})
\ee
where $S\{p\}$ is the Shannon entropy:
\be \label{Shannonentropy1}
S\{p\} = - \sum_{n} p_{n} \ln p_{n}
\ee
and $\a$ is a Lagrange multiplier.
The stationary point of \eqref{Maxentropy1} provides the equilibrium distribution constrained
by the target RDF and we will show that the knowledge of this function allows to introduce
a notion of interaction potential in the model system. This quantity will be identified
with the target potential thus providing a solution of the inverse problem.

\subsubsection{\label{LowDensity}Low density solution}

In the low density limit the general strategy previously described can be easily carried
out.
In order to evaluate the stationary point of the functional \eqref{Maxentropy1}
we perform an expansion of the Shannon entropy in correlation functions.
Leaving aside the ideal-gas contribution which does not depend on the
configurational degrees of freedom we have:
\be \label{Shannonentropy2}
S\{p\} = \sum_{n\geq 2} S^{(n)}
\ee
Formula \eqref{Shannonentropy2} provides an expansion of the excess entropy organized
in powers of the density and in the low density limit the whole series is dominated
by the two-body contribution $S^{\sst{(2)}}$.

The solution of the inverse problem is straightforward and proceeds in two steps.
First of all we maximize the two-body Shannon entropy assuming that the dynamics in
the model system is governed by an (unknown) pairwise additive potential $\p_{m}(r)$.
For pairwise additive interactions the configurational part of the internal energy can
be expressed as:
\be \label{internalEnergy1}
U = \frac{\rho}{2} \int d {\bf r} g(r) \p_{m}(r)
\ee
so the ME estimate of the two-body entropy functional subjected to the average value of
the internal energy is given by the stationary configuration of the functional:
\be \label{Maxentropy2}
{\cal F}\{g\} = S_{\cal P}^{\sst{(2)}} + \a \left(
\frac{\rho}{2} \int d {\bf r} g(r) \p_{m}(r) - U \right)
\ee
maximizing \eqref{Maxentropy2} and imposing the constraint \eqref{internalEnergy1}
together with the thermodynamic relation $\b = \partial S / \partial U$  provides
the solution:
\be \label{RDFlowdensity}
g(r) = e^{-\b\p_{m}(r)}
\ee
which is the ME estimate of the two-body equilibrium distribution for a system with
pairwise interactions \footnote{We have used the two-body entropy expression given by
\eqref{relativentropy2}; the same result can be obtained starting from the expression
\eqref{relativentropy1} but in this case a further Lagrange multiplier enforcing a
normalization condition on $g(r)$ has to be introduced.}.
We recognize the first order contribution in the cluster expansion of the RDF.

The second step is realized by imposing equation \eqref{relativentropy3} which allows to
evaluate the ME estimate of the interaction potential $\p_{m}(r)$ constrained by the target RDF.
Since the vanishing of the relative entropy implies the equality of the two RDFs we obtain:
\be \label{EffectivePotentialLowdensity}
\b \p_{m}(r) = - \ln g_{\sst{0}}(r)
\ee
which is the ME solution of the inverse problem at low density.

\subsubsection{\label{HighDensity}High density solution: a Monte Carlo approach}

The correlators expansion of the excess entropy \eqref{Shannonentropy2} for a high
density system contains, apart from the two-body contribution, all higher order terms.
Since these quantities are unknown a direct maximization procedure of the excess
entropy, like the one performed in the low density limit, is unfeasible.
However, if the interaction potential is pairwise additive, the RDF still
codifies all the information needed to the solution of the inverse problem.
This is a direct consequence of the Henderson theorem \cite{Henderson1974}: the RDF determines
the interaction potential up to a constant, so its knowledge sets the whole configurational
part of the phase space distribution function and all the higher order terms in the entropy
expansion are theoretically determined if the two-body contribution is given.
Anyhow, since the explicit computation of these terms would require the knowledge
of the interaction potential, a direct maximization procedure cannot be performed and a
different approach has to be adopted.

The general strategy to achieve the entropy maximization is to recur to a `Monte Carlo like'
(MC) approach in which the configuration space of the model system is sampled along a random path.
So, as in the standard Metropolis-Monte Carlo (MMC) algorithm, the dynamical evolution of the
system is defined by introducing a notion of trial configurations and a transition probability
between neighbour states.
We shall see that the stochastic nature of the MC dynamics together with a suitable choice of
the transition probability will allow to generate a path in the configuration space of the model
system which maximizes the excess entropy \eqref{Shannonentropy1} consistently with the relative
entropy constraint \eqref{relativentropy3}.

Let us define the building blocks of this procedure.
Assume that we have performed $s$ MC iterations. For each point of the path we compute a
local sampling of the PF and sum up these measurements in the global pair function \eqref{mdef1}.
Then we select a reference particle and compute a local sampling of the PF $n^{\sst{(1)}}$,
at the same time the particle is randomly moved and the new local sampling of the PF is stored
in the array $n^{\sst{(2)}}$. This procedure provides two different samplings of the global PF at
the level $s+1$:
\be \label{trialconf1}
m^{\sst{(1)}} = m + n^{\sst{(1)}} \qquad
m^{\sst{(2)}} = m + n^{\sst{(2)}}
\ee
the trial configuration $m^{\sst{(2)}}$ is accepted with a probability:
\be \label{TransiotionProbability1}
P_{m^{\sst{(1)}} \rightarrow m^{\sst{(2)}}} =
\min \big(1,f(m^{\sst{(1)}},m^{\sst{(2)}}) \big)
\ee
where $f$ is the transition probability which determines the stochastic evolution law.
The iteration of this procedure allows to generate the whole ensemble of configurations
of the model system.

Now we impose the constraint \eqref{relativentropy3}. To achieve this task we define the
transition probability by the requirement that the global PF \eqref{mdef1} built along
the path maximizes the likelihood function \eqref{logprobability1} with the reference pair
function $\mu$, defined in term of the target RDF via the relation:
\be \label{mushelldef1}
\mu_{i} = 4\pi\rho r_{i}^{2} g_{\sst{0}}(r_{i}) \d r
\ee
If we are able to impose this condition then equations \eqref{logprobability2} guarantees
that, asymptotically, the relative entropy between the model and target RDFs vanishes and
the constraint \eqref{relativentropy3} is satisfied.
For this purpose we try to guess a formula for the transition probability written in term
of a likelihood ratio:
\be \label{Likelihoodratio1}
f=e^{-\d \l} \, , \,\,\, {\rm where} \,\,\,\,
\d \l = \ln \frac{{\cal L}(m^{\sst{(1)}},\mu)}
{{\cal L}(m^{\sst{(2)}},\mu)}
\ee
so trial samples with a likelihood higher than $m^{\sst{(1)}}$ are automatically accepted,
otherwise they are accepted with a probability given by $f$.
For $s \gg 1$ we can make use of the Stirling approximation \eqref{logprobability1} for the
log-likelihood terms in \eqref{Likelihoodratio1}. Moreover, since the $n_{i}$ are of order
1 while the $m_{i}$ are of order $s$ we can expand in power of $s$ the logarithms appearing
in \eqref{logprobability1}. Performing this approximation to the first order in $1/s$
provides:
\be \label{Likelihoodratio2}
\d \l =
\sum_{i=1}^{J} \left(n_{i}^{\sst{(2)}} - n_{i}^{\sst{(1)}} \right)
\ln \frac{m_{i}}{s\mu_{i}}
\ee
this formula can be obtained starting from both the expressions for the log-likelihood
given in \eqref{logprobability1}, so the transition probability \eqref{Likelihoodratio2}
turns out to be invariant respect to the boundary condition imposed in the model system.

Equation \eqref{Likelihoodratio2} computes the difference among $n^{\sst{(1)}}$ and
$n^{\sst{(2)}}$ weighting each shell with a term:
\be\label{PIerror1}
e_{i}^{(s)} = \ln \frac{m_{i}}{s\mu_{i}}
\ee
that represents the `error' after $s$ iterations between the reference and the measured values
of the global PF. So $\d \l$ realizes a feedback in the model system, since it behaves as a controller
which selects the configurations in the model ensemble on the basis of the error \eqref{PIerror1}.
This controller operates only by considering the error in actual state $s$ and, adopting the common
language of the feedback control systems \cite{Abramovici2000}, we will call this quantity a
`proportional' controller.

The transition probability \eqref{Likelihoodratio2}, realized as a proportional controller, suffers
of a difficulty which is commonly encountered in many feedback controlled systems whenever the controller
is realized only through a proportional term: the presence of an offset between the measured process
variable and the target reference function.
Indeed a MC simulation built with this transition probability produces a model RDF which is a `biased'
reconstruction of the target one, so the formula guessed for $\d \l$ turns out to be inadeguate
to enforce a complete maximization of the likelihood function ${\cal L}(m,\mu)$.
A possible solution of this problem can be accomplished by realizing the control mechanism as a
proportional-integral controller (PI) \cite{Abramovici2000}. So we propose a modified expression for
$\d \l$ given by:
\be \label{PIoutput1}
\d \l =
\sum_{i=1}^{J} \left(n_{i}^{\sst{(2)}} - n_{i}^{\sst{(1)}} \right)
u_{i}^{(s)}
\ee
where $u_{i}$ is a function of the error \eqref{PIerror1} which depends on three different
contributions: a proportional term that determines the reaction to the current error,
an integral term which keeps into account the sum of all the former ones and a background
value which allows to include a priori knowledge on the system.
The output of the PI is given by a weighted sum of these three quantities:
\be \label{PIoutput2}
u_{i}^{(s)} = k^{(s)}_{p} e_{i}^{(s)} +
\sum_{\a=1}^{s} k^{(\a)}_{I} e_{i}^{(\a)}  + u_{i}^{\sst{(0)}}
\ee
where $k_{p}$ and $k_{I}$ are the ($s$ dependent) coefficients of the proportional and of
the integral terms.

A transition probability defined in term of the PI \eqref{PIoutput2} ensures that the model
RDF converges to its reference value. Furthermore the implementation of this controller
allows one to define an interaction potential in the model system.
In fact, as long as the measured PF converges to its reference value, the error \eqref{PIerror1}
goes to zero. In this limit the proportional term of \eqref{PIoutput2} becomes negligible and
the integral approaches to a constant finite value.
Formally we can define the model potential as the asymptotic limit of PI controller $(\b\equiv 1/k_{B}T)$:
\be \label{EffectivePotential1}
\b\p_{m}(r_{i}) = \lim_{s \rightarrow \infty} u_{i}^{(s)} =
\sum_{\a=1}^{\infty} k^{(\a)}_{I} e_{i}^{(\a)}
+ u_{i}^{\sst{(0)}}
\ee
So the MC dynamics built with the PI control system behaves as a constructive tool for
the computation of the model potential.
During a MC simulation the model system is subjected to a transient dynamical phase in which
the transition probability evolves during the path; as long as the path proceeds the PI
\emph{builds} the model potential \eqref{EffectivePotential1} and the transition probability
approaches to a stationary regime.
Once the equilibrium has been reached the system evolves according to a stationary transition
probability and behaves as a Markov chain, in which the potential is given by
\eqref{EffectivePotential1}.

\subsubsection{\label{ComputationPI}Computation of the PI coefficients}

Let us come back to the issue of the correct definition of the coefficients $k_{p}$ and $k_{I}$.
Usually the PI parameters are tuned with the aim to ensure a fast and stable convergence
of the measured process variable to its reference value.
In this case we propose a criterium, for fixing these parameters, which comes again from statistical
considerations. We observe that if the model system is sampled with the expected distribution
\eqref{poissondef1}, the global PF approaches to $s\mu_{i}$ as long as $s$ increases.
So we introduce the reduced variables $x_{i}$  defined by:
\be\label{reducedvariable1}
\frac{m_{i}}{s\mu_{i}} = 1 +
\frac{m_{i}-s\mu_{i}}{s\mu_{i}} = 1 + x_{i}
\ee
and we expand the distribution function \eqref{poissondef1} in series around $x_{i} = 0$.
Performing this expansion together with the usual Stirling approximation provides:
\be\label{poissondef2}
{\cal P}_{\mu}(m) \simeq \prod_{i=1}^{J}
\frac{1}{\sqrt{2\pi s\mu_{i}}} \,
e^{-\frac{1}{2}\frac{\left(m_{i}-s\mu_{i}\right)^2}{s\mu_{i}}}
\ee
so for large values of $s$ the global PF is distributed according to a product of $J$ gaussian
distributions \footnote{If we perform the same expansion starting from the multinomial distribution
\eqref{multinomdef1} we obtain a multivariate gaussian distribution in which the correlations between
different shells are imposed by the constraint \eqref{multinomconstraint1}. However, in the asymptotic
limit the off-diagonal elements of the covariance matrix become negligible and we obtain again the
result \eqref{poissondef2}.}.
Since the reduced variables $x_{i}$ are distributed according to a standard
normal distribution, the variable defined as:
\be\label{chisquare1}
\c^{2}_{(s)} \doteq \ \frac{1}{J} \sum_{i=1}^{J}
\frac{\left(m_{i}-s\mu_{i}\right)^2}{s\mu_{i}}
\ee
follows a $\c$-square distribution with $J$ degrees of freedom.

So we define the PI coefficients in order to implement the condition $\c^{2}=1$.
Enforcing this condition in the model system guarantees that the global PF has the correct
fluctuation around its average value and excludes spurious correlation among different shells.
This can be done by introducing a new PI which performs a dynamic control on
the coefficients $k_{p}$ and $k_{I}$, so we set:
\bea\label{chisquare2}
k_{p}^{(s)} &=& c_{\sst{1}}\left(\c^{2}_{(s)}-1\right) +
c_{\sst{2}}\sum_{\a=1}^{s} \left(\c^{2}_{(\a)}-1\right)
\nonumber \\
k_{I}^{(s)} &=& d_{\sst{1}}\left(\c^{2}_{(s)}-1\right) +
d_{\sst{2}}\sum_{\a=1}^{s} \left(\c^{2}_{(\a)}-1\right)
\eea
where $c_{\sst{1}},c_{\sst{2}},d_{\sst{1}},d_{\sst{2}}$ are the PI parameters.
Further details concerning the implementation
of this control mechanism will be given in section \ref{Applications}.

\section{\label{Applications}Applications}

In order to illustrate the features of the technique here proposed we have solved two
systems which have been widely analyzed in the literature concerning the inversion
methods \cite{Reatto1986,Almarza2003}: a simple Lennard-Jones fluid and a model for
liquid aluminum \cite{Dagens1975}.

We briefly describe the general strategy adopted in the analysis of both systems.
The target RDF has been evaluated recurring to a MMC simulation in the NVT ensemble.
The configuration space of the target system is a cubic volume of linear length $L$ with
$N_{p}$ point-like particle and the periodic boundary conditions together with the minimum
image convention have been adopted.
The target potential $\p(r)$ is truncated at $L/2$ and the system evolves starting from
an FCC lattice; after about $5\times 10^{2}$ MMC steps the energy of the system approaches
to a constant value and the system evolves around equilibrium.
Once at equilibrium a local sampling of the PF is performed for each configuration
and the average value of $\mu$ is built, then the target $g(r)$ is computed.
Due to the minimum image convention this method provides a reliable RDF up to the edge
value $r_{M}=L/2$. The error on the target RDF can be estimated by dividing the whole
simulation in blocks and by computing the standard deviation $\d g(r)$ between the blocks.

Once the $g(r)$ has been computed the inverse procedure for the determination
of the pair potential described in section \ref{Theory} can be applied.
The model system is realized exactly as the target one, so the configuration spaces of
the two systems are identical.
The PI coefficients are dynamically defined by equations \eqref{chisquare2} which
ensure the correct equilibrium fluctuation of the model RDF.
A direct analysis of the system response evidences that an optimal choice of the
parameters appearing in \eqref{chisquare2} is given by:
\bea\label{chisquare3}
k_{p}^{(s)} &=& \left(\c^{2}_{(s)}-1\right) + 1\times 10^{-3}
\sum_{\a=1}^{s} \left(\c^{2}_{(\a)}-1\right)
\nonumber \\
k_{I}^{(s)} &=& 5\times 10^{-3} k_{p}^{(s)}
\eea
where the ratio between $k_{p}$ and $k_{I}$ has been set to a constant value. This choice
guarantees a smooth convergence of the measured PF to the target reference value.
It is worth noting that performing a different choice (inside a range of values which does
not produce an oscillating behavior) has only the effect of changing the rate of convergence
of the model system but does not affect the convergence value.
Furthermore, the same set of parameters given by \eqref{chisquare3} have been used
both in the analysis of the Lennard-Jones fluid and of the liquid aluminum, providing
an equally good convergence independently of the details of the system.

We observe that the target RDF of both the systems under inspection exhibits a
hard core structure, i.e. $g(r)=0$ for $r<r_{\sst{0}}$.
This information can be imposed in the model by introducing a hard sphere (HS) background
potential, $u^{\sst{(0)}}= \infty$ for $r<r_{\sst{0}}$ and zero otherwise, which initializes
the PI controller \eqref{PIoutput2}. Due to this term any trial configuration containing
particle at a distance lower than $r_{\sst{0}}$ is automatically rejected.
Consistently with the background potential, we choose an equilibrium HS configuration as the
starting point for the MC path. Then the reverse procedure starts and the system evolves according
to the transition probability \eqref{PIoutput1}; after each iteration we compute the output of the
PI \eqref{PIoutput2} and the expression of the transition probability is updated.
Since the RDF of the starting configuration is noticeably different from the reference value,
the $\c^{2}$ is sensibly higher than 1 and the PI coefficients \eqref{chisquare3}
grow very fast; this phase is characterized by a highly non stationary dynamical evolution
of the transition probability \eqref{PIoutput1}.

In order to improve the convergence of the model potential it is convenient to split the simulation
into two phases.
So, when the $\c^{2}$ has reached a value quite close to 1 the actual configuration and the
final expression of the PI output are stored in a file and we stop the simulation.
Then these quantities are used as input values for the background potential and for
the initial configuration and we start the `refinement phase'.
Since the system is closer to equilibrium, the PI \eqref{chisquare3} works in a different
regime with respect to the previous phase; so the system evolves smoothly to equilibrium and
the transition probability approaches to its asymptotic value. This phase can be repeated many
times in order to obtain a better refinement of the model potential.

As a final check of the goodness of the results provided by this procedure we perform a standard
MMC simulation using the model potential and we compare the corresponding RDF with the target one.
If the difference of the two RDFs is not bigger than their intrinsical noise we conclude that the
model potential \eqref{EffectivePotential1} is equivalent to the target one and the reconstruction
procedure stops; otherwise further refinement phases could be needed.

\subsection{\label{LJ}Lennard-Jones potential}

The system is defined by a Lennard-Jones potential:
\be\label{LJpotential}
\p_{LJ}(r) = 4\eps \left[ \left(\frac{\s}{r}\right)^{12} -
\left(\frac{\s}{r}\right)^{6} \right]
\ee
with argon-like parameters $\s=3.405$ \AA\ and $\eps/k_{B}T=119.76$.
The MMC simulation for the determination of the target RDF is performed on a
system of 864 particle at the reduced density $\rho^{*}=\rho \s^{3}=0.84$ and
reduced temperature $T^{*}=k_{B}T/\epsilon=0.75$, near the triple point.
The $g(r_{i})$ has been evaluated up to $r^{*}=r/\s=5.0$ which corresponds to $L/2$;
the width of the shells for the measure of the $g(r_{i})$ was $\d r = 2.5\times 10^{-2}$
\AA\ and the number of measured points was 686.
We performed $2\times 10^4$ cycles after equilibration.
The experimental error on the RDF was estimated by computing the standard deviation
$\d g(r_{i})$ between 50 blocks of $4\times 10^2$ cycles each. The largest value for
$\d g(r_{i})$ was about $2\times 10^{-2}$ with an average value of $7\times 10^{-3}$.

\begin{figure}
\includegraphics[width=9cm,height=8cm]{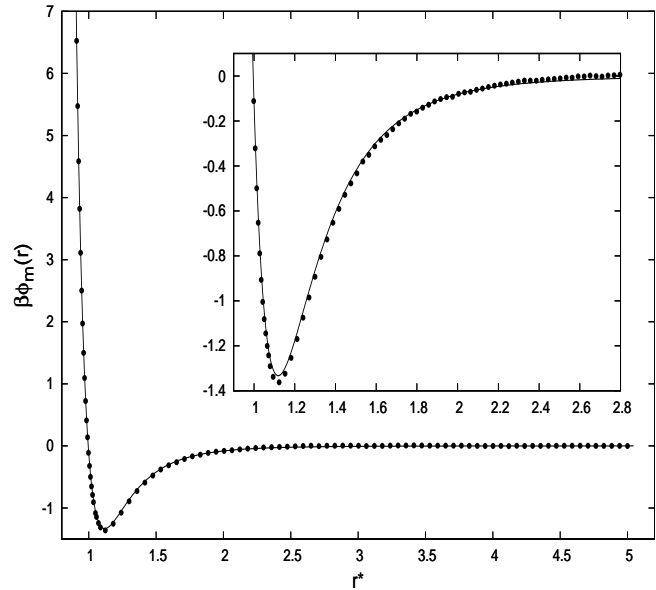}
\caption{\label{LJresult}Results for the Lennard-Jones system. The target
potential (continuous line) and the model potential (filled circles)
are plotted.}
\end{figure}

The complete inverse simulation procedure took $2.4\times 10^4$ iteration.
A first phase of $6\times 10^3$ steps was performed starting from the FCC lattice and then
the refinement phase was repeated three times for $6\times 10^3$ steps each.
The result for the interaction potential is reported in Fig. \ref{LJresult},
the maximum difference between the model potential and the Lennard-Jones
reference one was less than $5\times 10^{-2}$ with an average value of $1\times 10^{-2}$.
The average difference between the model and the target RDFs was equal to $4\times 10^{-3}$;
this value is inside the average noise of the RDF, so the model potential of Fig. \ref{LJresult}
can be considered identical to the Lennard-Jones one.

\subsection{\label{Al}Model potential of aluminum}

The system is defined by a model potential for liquid aluminum \cite{Dagens1975}.
The MMC simulation for the determination of the target RDF was performed on a system of
864 particle at the density $\rho = 0.0527$ \AA${}^{-3}$ and $T=1051$ $K$.
The $g(r_{i})$ has been evaluated up to $r=12.70$ \AA\ which corresponds to $L/2$;
the width of the shells for the measure of the $g(r_{i})$ was $\d r = 2.5\times 10^{-2}$
\AA\ and the number of measured points was 508.
We performed $2\times 10^4$ cycles after equilibration. The experimental error on the RDF
was estimated by computing the standard deviation $\d g(r_{i})$ between 50 blocks of $4\times 10^2$
cycles each. The largest value for $\d g(r_{i})$ was about $2\times 10^{-2}$ with an average value
of $6\times 10^{-3}$.

\begin{figure}
\includegraphics[width=9cm,height=8cm]{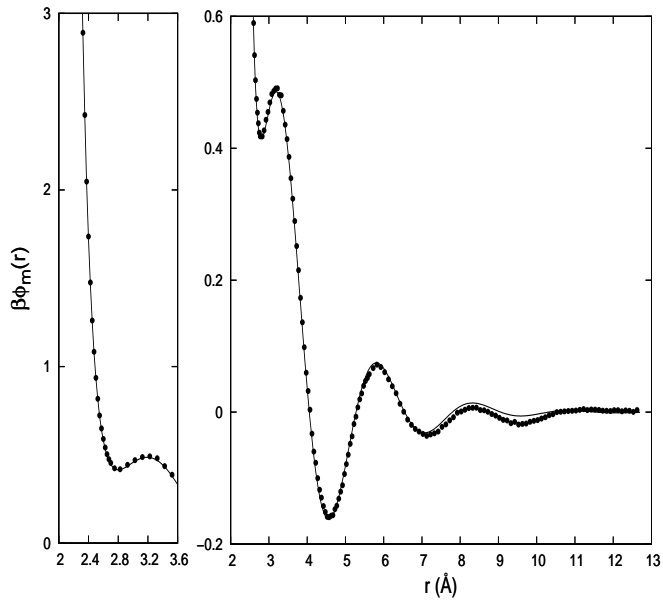}
\caption{\label{Alresult}Results for aluminum. The aluminum
potential (continuous line) and the model potential (filled circles)
are plotted.}
\end{figure}

The inverse simulation procedure took $2.6\times 10^4$ iteration.
A first phase of $6\times 10^3$ steps was performed starting from the FCC lattice and then the
refinement phase was repeated twice for $6\times 10^3$ steps each and once for $8\times 10^3$ steps.
The result for the interaction potential is reported in Fig. \ref{Alresult},
the maximum difference between the model potential and the Al model reference value was less
than $2\times 10^{-2}$ with an average value of $7\times 10^{-3}$.
Even in this case the average difference between the model and the target RDFs
is inside the typical noise of the RDF.
Analyzing Fig. \ref{Alresult} we observe a difference between the target and model potential
of the order of $1\times 10^{-2}$ in the range from 7 to 11 \AA . This error is due to a correlated statistical
fluctuation in the reconstruction procedure and can be further reduced by increasing the information content
in the target RDF used as input.

\section{\label{Discussion}Discussion and conclusions}

The method presented so far supplies an accurate solution to the issue of determining
the interaction potential from the radial distribution function.
This technique bases its theoretical support on the maximum entropy principle of information
theory which provides a general tool for the statistical inference on the basis of partial
knowledge.
The method is formally summarized by equation \eqref{Maxentropy1} which describes the
maximization of the configurational entropy ($S$ term) constrained by the information codified in the
target system ($K$ term).
The ME solution is sought inside a Monte Carlo scheme where the maximization of configurational
entropy is realized through the MC random displacements and the acceptance criterion for the trial configurations
is built consistently with the physical input provided by the target RDF.
The potential emerges as the asymptotic expression of the transition probability and, for pairwise
potentials, it reproduces completely the interactions of the target system.
This method fulfills some nice properties that, in our opinion, make it a valid tool for the extraction
of potential. Actually the expression of the transition probability \eqref{PIoutput1} is motivated only
by the constraint \eqref{relativentropy3} and does not rely on any ulterior hypothesis concerning the
physical nature of the target system, so we expect that the general strategy depicted in the present
paper could be of wide applicability. Nevertheless, the convergence of the model potential is ensured
by a feedback control mechanism and the coefficients of \eqref{PIoutput2} are tuned by an independent
PI which operates a control on the fluctuation of the model RDF around the target reference value.
This further controller avoids spurious correlations in the model RDF and guarantees that no information,
besides the one codified in the target RDF, is transferred to the model during the simulation.

Results of section \ref{Applications} show that the extracted potential \eqref{EffectivePotential1}
accurately reproduces the original pair interaction both for the Lennard-Jones fluid and for the liquid
aluminum model.
A comparison between these results and the ones presented in \cite{Reatto1986} and \cite{Almarza2003}
evidences a very satisfactory accuracy, despite a modest computational effort.
This level of agreement turns out to be highly remarkable since the systems lie in the high density region
of the state space where it is expected that the RDF should be quite insensitive to the details of the interaction;
moreover the aluminum potential exhibits well defined oscillations even at short distances, where $\b\p(r)$ is
still positive.
As a further control we have verified that the method provides the correct results in a different region of the
$(\rho,T)$ plane; so the procedure described in section \ref{Applications} has been repeated for a Lennard-Jones
fluid at $\rho^{*}=0.5$, $T^{*}=1$. As expected, the interaction potential approaches the correct result with
a convergence rate even faster than in the high density case (about $1\times 10^4$
steps were needed to obtain an accuracy comparable with the result of Fig. \ref{LJresult}).
This analysis indicates that our procedure for the solution of the inverse problem provides reliable results
independently both from the density of the system and the shape of the potential under inspection, so it fulfills
the requirements of a `satisfactory inversion scheme' as stated in \cite{Reatto1988}.

The interpretation of the transition probability as a feedback controller represents a key point for the
accomplishment of the solution discussed in the present paper. Actually, the adoption of this point of view
motivates the introduction of the integral term and gives rise to the model interaction potential \eqref{EffectivePotential1}.
We want to point out that this is not the only way to impose the constraint \eqref{relativentropy3}.
For example, the offset between the target reference function $\mu$  and the model global PF can be made
null by using a proportional controller with an infinite value of the coefficient $k_{p}$.
Pursuing this approach leads to the `pure minimization methods' \cite{Cilloco1993} and \cite{daSilva1998} in
which only trial configurations with a higher likelihood function (or with a lower $\chi^{2}$ in the language
of \cite{daSilva1998}) are accepted.
The drawback of this approach is that, due to the lacking of the integral term, the interaction potential cannot
be directly computed.

We conclude our discussion with some comments concerning the extension of this procedure to other systems than
the simple monoatomic fluid analyzed in the present paper.
The method is based on ME principle which holds for any system at equilibrium.
For simple fluids a $K$ term realized as the relative entropy (\ref{relativentropy1},\ref{relativentropy2})
between the RDFs is able to constrain the whole configurational part of the probability distribution function
in the model system. The information closed loop realized by the PI controller \eqref{PIoutput2} then allows one
to determine completely the interaction potential.
Conversely, if we are dealing with more complex systems, that contain further degrees of freedom beside the position
of the center of mass of the atoms, a ME solution is always possible, which will correspond to an effective potential.
If, however, the complete target potential is sought, then it is necessary to match the relevant degrees of
freedom of the systems with further involvement of information; for instance the experimental three body
correlation function and the inclusion of higher order terms in the definition of $K$ would be necessary if a
three body interaction is present.
As a final remark, we point out that this inversion technique has been discussed assuming that the RDF of the target
system is given.
However, since experimental data are expressed in term of the structure factor, a preliminary transformation
to the real space RDF has to be performed in order to extract the interaction potential of a real system.
This procedure may be hampered by the limited range of the structure factor or by the unsatisfactory $k$-resolution
so, again, the use of the ME methods could reveal a useful tool to overcome those problems in optimal way.

\begin{acknowledgments}
The authors wish to thank Luciana Silvestri for the linguistic revision of the manuscript.
\end{acknowledgments}

\bibliography{dalessandroReference-final}

\begin{thebibliography}{24}%
\makeatletter
\providecommand \@ifxundefined [1]{%
 \@ifx{#1\undefined}
}%
\providecommand \@ifnum [1]{%
 \ifnum #1\expandafter \@firstoftwo
 \else \expandafter \@secondoftwo
 \fi
}%
\providecommand \@ifx [1]{%
 \ifx #1\expandafter \@firstoftwo
 \else \expandafter \@secondoftwo
 \fi
}%
\providecommand \natexlab [1]{#1}%
\providecommand \enquote  [1]{``#1''}%
\providecommand \bibnamefont  [1]{#1}%
\providecommand \bibfnamefont [1]{#1}%
\providecommand \citenamefont [1]{#1}%
\providecommand \href@noop [0]{\@secondoftwo}%
\providecommand \href [0]{\begingroup \@sanitize@url \@href}%
\providecommand \@href[1]{\@@startlink{#1}\@@href}%
\providecommand \@@href[1]{\endgroup#1\@@endlink}%
\providecommand \@sanitize@url [0]{\catcode `\\12\catcode `\$12\catcode
  `\&12\catcode `\#12\catcode `\^12\catcode `\_12\catcode `\%12\relax}%
\providecommand \@@startlink[1]{}%
\providecommand \@@endlink[0]{}%
\providecommand \url  [0]{\begingroup\@sanitize@url \@url }%
\providecommand \@url [1]{\endgroup\@href {#1}{\urlprefix }}%
\providecommand \urlprefix  [0]{URL }%
\providecommand \Eprint [0]{\href }%
\@ifxundefined \urlstyle {%
  \providecommand \doi  [0]{\begingroup \@sanitize@url \@doi}%
  \providecommand \@doi [1]{\endgroup \@@startlink {\doibase
  #1}doi:\discretionary {}{}{}#1\@@endlink }%
}{%
  \providecommand \doi  [0]{doi:\discretionary{}{}{}\begingroup
  \urlstyle{rm}\Url }%
}%
\providecommand \doibase [0]{http://dx.doi.org/}%
\providecommand \Doi [0]{\begingroup \@sanitize@url \@Doi }%
\providecommand \@Doi  [1]{\endgroup\@@startlink{\doibase#1}\@@Doi}%
\providecommand \@@Doi [1]{#1\@@endlink}%
\providecommand \selectlanguage [0]{\@gobble}%
\providecommand \bibinfo  [0]{\@secondoftwo}%
\providecommand \bibfield  [0]{\@secondoftwo}%
\providecommand \translation [1]{[#1]}%
\providecommand \BibitemOpen [0]{}%
\providecommand \bibitemStop [0]{}%
\providecommand \bibitemNoStop [0]{.\EOS\space}%
\providecommand \EOS [0]{\spacefactor3000\relax}%
\providecommand \BibitemShut  [1]{\csname bibitem#1\endcsname}%
\bibitem [{\citenamefont {Henderson}(1974)}]{Henderson1974}%
  \BibitemOpen
  \bibfield  {author} {\bibinfo {author} {\bibfnamefont {R.~L.}\ \bibnamefont
  {Henderson}},\ }\href@noop {} {\bibfield  {journal} {\bibinfo  {journal}
  {Phys. Lett.},\ }\textbf {\bibinfo {volume} {49A}},\ \bibinfo {pages} {197}
  (\bibinfo {year} {1974})}\BibitemShut {NoStop}%
\bibitem [{\citenamefont {Reatto}(1988)}]{Reatto1988}%
  \BibitemOpen
  \bibfield  {author} {\bibinfo {author} {\bibfnamefont {L.}~\bibnamefont
  {Reatto}},\ }\href@noop {} {\bibfield  {journal} {\bibinfo  {journal}
  {Philosophical Magazine A},\ }\textbf {\bibinfo {volume} {58}},\ \bibinfo
  {pages} {37} (\bibinfo {year} {1988})}\BibitemShut {NoStop}%
\bibitem [{\citenamefont {Schommers}(1983)}]{Schommers1983}%
  \BibitemOpen
  \bibfield  {author} {\bibinfo {author} {\bibfnamefont {W.}~\bibnamefont
  {Schommers}},\ }\href@noop {} {\bibfield  {journal} {\bibinfo  {journal}
  {Phys. Rev. A},\ }\textbf {\bibinfo {volume} {28}},\ \bibinfo {pages} {3599}
  (\bibinfo {year} {1983})}\BibitemShut {NoStop}%
\bibitem [{\citenamefont {Reatto}\ \emph {et~al.}(1986)\citenamefont {Reatto},
  \citenamefont {Levesque},\ and\ \citenamefont {Weis}}]{Reatto1986}%
  \BibitemOpen
  \bibfield  {author} {\bibinfo {author} {\bibfnamefont {L.}~\bibnamefont
  {Reatto}}, \bibinfo {author} {\bibfnamefont {D.}~\bibnamefont {Levesque}}, \
  and\ \bibinfo {author} {\bibfnamefont {J.~J.}\ \bibnamefont {Weis}},\
  }\href@noop {} {\bibfield  {journal} {\bibinfo  {journal} {Phys. Rev. A},\
  }\textbf {\bibinfo {volume} {33}},\ \bibinfo {pages} {3451} (\bibinfo {year}
  {1986})}\BibitemShut {NoStop}%
\bibitem [{\citenamefont {Soper}(1996)}]{Soper1996}%
  \BibitemOpen
  \bibfield  {author} {\bibinfo {author} {\bibfnamefont {A.~K.}\ \bibnamefont
  {Soper}},\ }\href@noop {} {\bibfield  {journal} {\bibinfo  {journal} {J.
  Chem. Phys.},\ }\textbf {\bibinfo {volume} {202}},\ \bibinfo {pages} {295}
  (\bibinfo {year} {1996})}\BibitemShut {NoStop}%
\bibitem [{\citenamefont {Soper}(2001)}]{Soper2001}%
  \BibitemOpen
  \bibfield  {author} {\bibinfo {author} {\bibfnamefont {A.~K.}\ \bibnamefont
  {Soper}},\ }\href@noop {} {\bibfield  {journal} {\bibinfo  {journal} {Mol.
  Phys.},\ }\textbf {\bibinfo {volume} {99}},\ \bibinfo {pages} {1503}
  (\bibinfo {year} {2001})}\BibitemShut {NoStop}%
\bibitem [{\citenamefont {Lyubartsev}\ and\ \citenamefont
  {Laaksonen}(1995)}]{Lyubartsev1995}%
  \BibitemOpen
  \bibfield  {author} {\bibinfo {author} {\bibfnamefont {A.~P.}\ \bibnamefont
  {Lyubartsev}}\ and\ \bibinfo {author} {\bibfnamefont {A.}~\bibnamefont
  {Laaksonen}},\ }\Doi {10.1103/PhysRevE.52.3730} {\bibfield  {journal}
  {\bibinfo  {journal} {Phys. Rev. E},\ }\textbf {\bibinfo {volume} {52}},\
  \bibinfo {pages} {3730} (\bibinfo {year} {1995})}\BibitemShut {NoStop}%
\bibitem [{\citenamefont {McGreevy}\ and\ \citenamefont
  {Pusztai}(1988)}]{McGreevy1988}%
  \BibitemOpen
  \bibfield  {author} {\bibinfo {author} {\bibfnamefont {R.~L.}\ \bibnamefont
  {McGreevy}}\ and\ \bibinfo {author} {\bibfnamefont {L.}~\bibnamefont
  {Pusztai}},\ }\href@noop {} {\bibfield  {journal} {\bibinfo  {journal} {Mol.
  Simul.},\ }\textbf {\bibinfo {volume} {1}},\ \bibinfo {pages} {359} (\bibinfo
  {year} {1988})}\BibitemShut {NoStop}%
\bibitem [{\citenamefont {Cilloco}(1993)}]{Cilloco1993}%
  \BibitemOpen
  \bibfield  {author} {\bibinfo {author} {\bibfnamefont {F.}~\bibnamefont
  {Cilloco}},\ }\href@noop {} {\bibfield  {journal} {\bibinfo  {journal} {J. of
  Mol. Struct.},\ \bibinfo {pages} {253}} (\bibinfo {year} {1993})}\BibitemShut
  {NoStop}%
\bibitem [{\citenamefont {da~Silva}\ \emph {et~al.}(1998)\citenamefont
  {da~Silva}, \citenamefont {Svensson}, \citenamefont {Akesson},\ and\
  \citenamefont {Jonsson}}]{daSilva1998}%
  \BibitemOpen
  \bibfield  {author} {\bibinfo {author} {\bibfnamefont {F.~L.~B.}\
  \bibnamefont {da~Silva}}, \bibinfo {author} {\bibfnamefont {B.}~\bibnamefont
  {Svensson}}, \bibinfo {author} {\bibfnamefont {T.}~\bibnamefont {Akesson}}, \
  and\ \bibinfo {author} {\bibfnamefont {B.}~\bibnamefont {Jonsson}},\
  }\href@noop {} {\bibfield  {journal} {\bibinfo  {journal} {J. of Chem.
  Phys.},\ }\textbf {\bibinfo {volume} {109}},\ \bibinfo {pages} {2624}
  (\bibinfo {year} {1998})}\BibitemShut {NoStop}%
\bibitem [{\citenamefont {Almarza}\ and\ \citenamefont
  {Lomba}(2003)}]{Almarza2003}%
  \BibitemOpen
  \bibfield  {author} {\bibinfo {author} {\bibfnamefont {N.~G.}\ \bibnamefont
  {Almarza}}\ and\ \bibinfo {author} {\bibfnamefont {E.}~\bibnamefont
  {Lomba}},\ }\Doi {10.1103/PhysRevE.68.011202} {\bibfield  {journal} {\bibinfo
   {journal} {Phys. Rev. E},\ }\textbf {\bibinfo {volume} {68}},\ \bibinfo
  {pages} {011202} (\bibinfo {year} {2003})}\BibitemShut {NoStop}%
\bibitem [{\citenamefont {Almarza}\ \emph {et~al.}(2004)\citenamefont
  {Almarza}, \citenamefont {Lomba},\ and\ \citenamefont
  {Molina}}]{Almarza2004}%
  \BibitemOpen
  \bibfield  {author} {\bibinfo {author} {\bibfnamefont {N.~G.}\ \bibnamefont
  {Almarza}}, \bibinfo {author} {\bibfnamefont {E.}~\bibnamefont {Lomba}}, \
  and\ \bibinfo {author} {\bibfnamefont {D.}~\bibnamefont {Molina}},\ }\Doi
  {10.1103/PhysRevE.70.021203} {\bibfield  {journal} {\bibinfo  {journal}
  {Phys. Rev. E},\ }\textbf {\bibinfo {volume} {70}},\ \bibinfo {pages}
  {021203} (\bibinfo {year} {2004})}\BibitemShut {NoStop}%
\bibitem [{\citenamefont {Jaynes}(1957)}]{Jaynes1957}%
  \BibitemOpen
  \bibfield  {author} {\bibinfo {author} {\bibfnamefont {E.~T.}\ \bibnamefont
  {Jaynes}},\ }\href@noop {} {\bibfield  {journal} {\bibinfo  {journal} {Phys.
  Rev.},\ }\textbf {\bibinfo {volume} {106}},\ \bibinfo {pages} {620} (\bibinfo
  {year} {1957})}\BibitemShut {NoStop}%
\bibitem [{Note1()}]{Note1}%
  \BibitemOpen
  \bibinfo {note} {The configuration space of the system is assumed large
  enough so that for each of the sampled elements the sphere of radius $r_{M}$
  does not cross the boundary of the system.}\BibitemShut {Stop}%
\bibitem [{\citenamefont {Baker}\ and\ \citenamefont
  {Cousins}(1984)}]{Baker1984}%
  \BibitemOpen
  \bibfield  {author} {\bibinfo {author} {\bibfnamefont {S.}~\bibnamefont
  {Baker}}\ and\ \bibinfo {author} {\bibfnamefont {R.~D.}\ \bibnamefont
  {Cousins}},\ }\Doi {DOI: 10.1016/0167-5087(84)90016-4} {\bibfield  {journal}
  {\bibinfo  {journal} {Nucl. Instr. and Meth. in Phys. Res.},\ }\textbf
  {\bibinfo {volume} {221}},\ \bibinfo {pages} {437} (\bibinfo {year}
  {1984})}\BibitemShut {NoStop}%
\bibitem [{\citenamefont {Kullback}\ and\ \citenamefont
  {Leibler}(1951)}]{Kullback1951}%
  \BibitemOpen
  \bibfield  {author} {\bibinfo {author} {\bibfnamefont {S.}~\bibnamefont
  {Kullback}}\ and\ \bibinfo {author} {\bibfnamefont {R.~A.}\ \bibnamefont
  {Leibler}},\ }\href@noop {} {\bibfield  {journal} {\bibinfo  {journal}
  {Annals Math. Stat.},\ }\textbf {\bibinfo {volume} {22}},\ \bibinfo {pages}
  {79} (\bibinfo {year} {1951})}\BibitemShut {NoStop}%
\bibitem [{\citenamefont {Shannon}(1948)}]{Shannon1948}%
  \BibitemOpen
  \bibfield  {author} {\bibinfo {author} {\bibfnamefont {C.~E.}\ \bibnamefont
  {Shannon}},\ }\href@noop {} {\bibfield  {journal} {\bibinfo  {journal} {Bell
  Sys. Tech. J.},\ }\textbf {\bibinfo {volume} {27}},\ \bibinfo {pages} {379}
  (\bibinfo {year} {1948})}\BibitemShut {NoStop}%
\bibitem [{\citenamefont {Green}(1952)}]{Green1952}%
  \BibitemOpen
  \bibfield  {author} {\bibinfo {author} {\bibfnamefont {H.~S.}\ \bibnamefont
  {Green}},\ }\href@noop {} {\emph {\bibinfo {title} {The Molecular Theory of
  Fluids}}}\ (\bibinfo  {publisher} {North-Holland, Amsterdam},\ \bibinfo
  {year} {1952})\BibitemShut {NoStop}%
\bibitem [{\citenamefont {Nettleton}\ and\ \citenamefont
  {Green}(1958)}]{Nettleton1958}%
  \BibitemOpen
  \bibfield  {author} {\bibinfo {author} {\bibfnamefont {R.~E.}\ \bibnamefont
  {Nettleton}}\ and\ \bibinfo {author} {\bibfnamefont {M.~S.}\ \bibnamefont
  {Green}},\ }\href@noop {} {\bibfield  {journal} {\bibinfo  {journal} {J.
  Chem. Phys.},\ }\textbf {\bibinfo {volume} {29}},\ \bibinfo {pages} {1365}
  (\bibinfo {year} {1958})}\BibitemShut {NoStop}%
\bibitem [{\citenamefont {Ravech\'e}(1971)}]{Raveche1971}%
  \BibitemOpen
  \bibfield  {author} {\bibinfo {author} {\bibfnamefont {H.~J.}\ \bibnamefont
  {Ravech\'e}},\ }\href@noop {} {\bibfield  {journal} {\bibinfo  {journal} {J.
  Chem. Phys.},\ }\textbf {\bibinfo {volume} {55}},\ \bibinfo {pages} {2242}
  (\bibinfo {year} {1971})}\BibitemShut {NoStop}%
\bibitem [{Note2()}]{Note2}%
  \BibitemOpen
  \bibinfo {note} {We have used the two-body entropy expression given by
  \protect \textup {\hbox {\mathsurround \z@ \protect \normalfont
  (\ignorespaces \ref {relativentropy2}\unskip \@@italiccorr )}}; the same
  result can be obtained starting from the expression \protect \textup {\hbox
  {\mathsurround \z@ \protect \normalfont (\ignorespaces \ref
  {relativentropy1}\unskip \@@italiccorr )}} but in this case a further
  Lagrange multiplier enforcing a normalization condition on $g(r)$ has to be
  introduced.}\BibitemShut {Stop}%
\bibitem [{\citenamefont {Abramovici}\ and\ \citenamefont
  {Chapsky}(2000)}]{Abramovici2000}%
  \BibitemOpen
  \bibfield  {author} {\bibinfo {author} {\bibfnamefont {A.}~\bibnamefont
  {Abramovici}}\ and\ \bibinfo {author} {\bibfnamefont {J.}~\bibnamefont
  {Chapsky}},\ }\href@noop {} {\emph {\bibinfo {title} {Feedback control
  systems: a fast-track guide for scientists and engineers}}}\ (\bibinfo
  {publisher} {Springer, New York},\ \bibinfo {year} {2000})\ p.\ \bibinfo
  {pages} {181}\BibitemShut {NoStop}%
\bibitem [{Note3()}]{Note3}%
  \BibitemOpen
  \bibinfo {note} {If we perform the same expansion starting from the
  multinomial distribution \protect \textup {\hbox {\mathsurround \z@ \protect
  \normalfont (\ignorespaces \ref {multinomdef1}\unskip \@@italiccorr )}} we
  obtain a multivariate gaussian distribution in which the correlations between
  different shells are imposed by the constraint \protect \textup {\hbox
  {\mathsurround \z@ \protect \normalfont (\ignorespaces \ref
  {multinomconstraint1}\unskip \@@italiccorr )}}. However, in the asymptotic
  limit the off-diagonal elements of the covariance matrix become negligible
  and we obtain again the result \protect \textup {\hbox {\mathsurround \z@
  \protect \normalfont (\ignorespaces \ref {poissondef2}\unskip \@@italiccorr
  )}}.}\BibitemShut {Stop}%
\bibitem [{\citenamefont {Dagens}\ \emph {et~al.}(1975)\citenamefont {Dagens},
  \citenamefont {Rasolt},\ and\ \citenamefont {Taylor}}]{Dagens1975}%
  \BibitemOpen
  \bibfield  {author} {\bibinfo {author} {\bibfnamefont {L.}~\bibnamefont
  {Dagens}}, \bibinfo {author} {\bibfnamefont {M.}~\bibnamefont {Rasolt}}, \
  and\ \bibinfo {author} {\bibfnamefont {R.}~\bibnamefont {Taylor}},\ }\Doi
  {10.1103/PhysRevB.11.2726} {\bibfield  {journal} {\bibinfo  {journal} {Phys.
  Rev. B},\ }\textbf {\bibinfo {volume} {11}},\ \bibinfo {pages} {2726}
  (\bibinfo {year} {1975})}\BibitemShut {NoStop}%
\end{thebibliography}%

\end{document}